\documentstyle[12pt,psfig]{article}

\renewcommand{\*}{ \hspace{-6pt}&=&\hspace{-6pt} }

\def\IR{{\rm I\!R}}
\def\H{{\rm H}}

\def\Tr{{\rm Tr}}

\def\Z{{\bf Z}}
\begin{document}

\newpage
\bigskip
\hskip 5in\vbox{\baselineskip12pt
\hbox{DTP/00/29}
\hbox{hep-th/0004068}}
{\flushleft\vskip-1.35cm\vbox{\psfig{figure=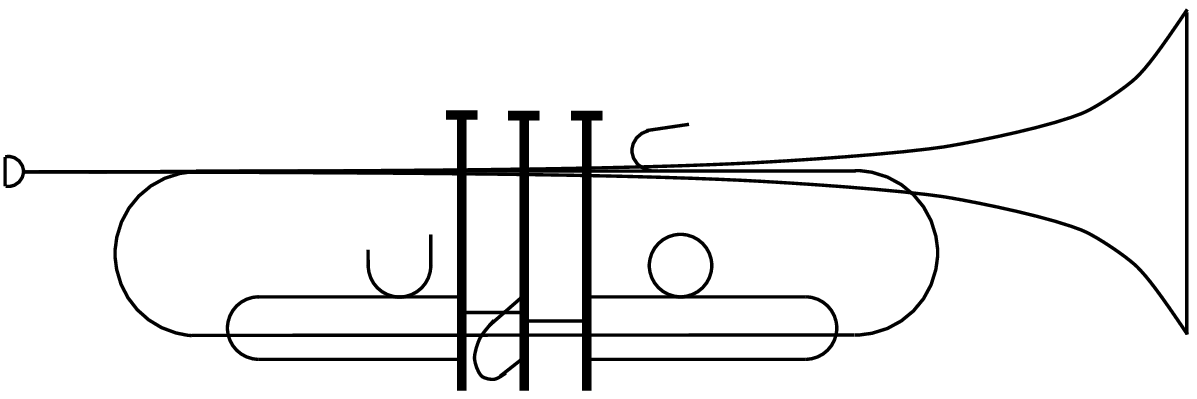,height=0.45in}}}

\bigskip
\bigskip
\bigskip
\bigskip

\centerline{\large \bf Enhan\c{c}ons, Fuzzy Spheres and Multi--Monopoles}
\bigskip\bigskip
\bigskip
\bigskip

\centerline{{\bf Clifford V. Johnson\footnote{c.v.johnson@durham.ac.uk}}}

\bigskip
\bigskip
\bigskip
\centerline{Centre for Particle Theory}
\centerline{Department of Mathematical Sciences}
\centerline{University of Durham, Durham DH1 3LE}
\centerline{England, U.K.}

\bigskip
\bigskip

\begin{abstract}
\noindent \baselineskip=16pt 
\smallskip

We study the ``enhan\c con'', a spherical hypersurface apparently made
of D--branes, which arises in string theory studies of large $N$
$SU(N)$ pure gauge theories with eight supercharges.  When
the gauge theory is 2+1 dimensional, the enhan\c con is an $S^2$. A
relation to charge~$N$ BPS multi--monopoles is exploited to uncover
many of its detailed properties. It is simply a spherical slice
through an Atiyah--Hitchin--like submanifold of the charge $N$ BPS
monopole moduli space. In the form of Nahm data, it is built from the
$N$ dimensional irreducible representation of $SU(2)$. In this sense
the enhan\c con is a non--commutative sphere, reminiscent of the
spherical ``dielectric'' branes of Myers. 
\end{abstract}
\newpage \baselineskip=18pt \setcounter{footnote}{0}


\section{Origins of the Enhan\c con}
Consider compactifying ten dimensional type~II string theory on the
four dimensional K3 surface of volume $V$. This gives a six
dimensional theory with ${\cal N}{=}2$ supersymmetry; in other words,
sixteen supercharges.  Consider further wrapping $N$
D$(p{+}4)$--branes on the $K3$.  Then there is an effective
$p$--dimensional extended object in the six dimensions.  It is in fact
a BPS solution, and there are eight supercharges preserved by this
situation.  Furthermore, there is an $SU(N)$ pure gauge theory with
eight supercharges on the $(p{+}1)$--dimensional world volume of the
BPS soliton. This soliton has a description as a bound state of the
wrapped brane and a negatively charged D$p$--brane\cite{jpp}. We shall
refer to this as the D$(p{+}4)$--D$p^*$ system, where the asterisk
($^*$) is to remind us that this is not an ordinary D$p$ brane, since
that would be an instanton.

Let us focus on $p=2$, hence studying type~IIA.  The supergravity
theory contains twenty--four $U(1)$'s coming from the various R-R
potentials in the theory. Of these, twenty--two come from wrapping the
two--form on the 19+3 two--cycles of $K3$.  The remaining two are
special $U(1)$'s for our purposes: One of them arises from wrapping
the five--form entirely on $K3$, while the final one is simply the
plain one--form already present in the uncompactified theory.
 
In fact, the BPS soliton is actually a monopole of one of the six
dimensional $U(1)$'s. It is obvious which $U(1)$ this is; the diagonal
combination of the two special ones we mentioned above.  Actually, we
can simply ignore the 2 spatial directions in which the soliton is
extended and see that the monopole sector (recall that it is also
coupled to gravity) is nothing more than the usual problem of
monopoles\cite{thooftpolyakov,poles} in a 3+1 dimensional gauge theory
with an adjoint Higgs. The first order ``Bogomolnyi''
equations\cite{bogo} are:
\begin{eqnarray}
&&B_i\equiv{1\over2}\epsilon_{ijk}F_{jk}=D_i\H\ ,\quad{\rm with}\nonumber\\
&&\quad F_{ij}=\partial_iA_j-\partial_jA_i+[A_i,A_j]; \quad \quad D_i\H=\partial_i\H+[A_i,\H]\ ,
\label{bogomolnyi}
\end{eqnarray}
with gauge invariance $(g({\bf x})\in SU(2))$:
\begin{equation}
A_i\to g^{-1}A_ig+g^{-1}\partial_ig;\quad \H\to g^{-1}\H g\ .
\end{equation}
Static, finite energy monopole solutions satisfy
\begin{equation}
\|\H({\bf x})\|\equiv{1\over2}\Tr \left[\H^*\H\right]\to H\  \quad {\rm as}\quad r\to\infty\ ,
\end{equation}
where ${\bf x}=(x_1,x_2,x_3)$ and $r^2=x_1^2+x_2^2+x_3^2$.  The
$SU(2)$ is spontaneously broken to $U(1)$, by the Higgs vacuum
expectation value (``vev'') $H$, whose magnetic charge the monopoles
carry. For orientation, and for later use, the explicitly known fields
of the one monopole solution is\cite{onemonopole}:
\begin{eqnarray}
\H(r)={1\over r}\left(\coth{r}-{1\over r}\right){\rm i}\sigma_ix_i\ ;\quad
A_i(r)={1\over r}\left({1\over\sinh r}-{1\over r}\right){\rm i}\epsilon_{ijk}\sigma_jx_k\ ,
\label{one}
\end{eqnarray}
where ${\bf x}=(0,0,r)$ and
\begin{equation}
\sigma_1=\pmatrix{0&1\cr1&0}\ ;\quad \sigma_2=\pmatrix{\phantom{-}0&{\rm i}\cr-{\rm i}&0}\ ;\quad 
\sigma_3=\pmatrix{1&\phantom{-}0\cr0&-1}\ .
\end{equation}
Note that it is spherically symmetric and has been normalized (for
later use) such that $H\to1-r^{-1}+\ldots$, as $r\to\infty$, with unit
magnetic charge.

Where did the $SU(2)$ come from? When the $K3$'s volume reaches the
value $V_*{\equiv}(2\pi\sqrt{\alpha^\prime})^4$, our $U(1)$ is
enhanced to $SU(2)$. This is a stringy phenomenon which has no
description in supergravity, since (for example) the W--bosons for
this $SU(2)$ are made of wrapped D4--branes.

This is an interesting system to use to study\cite{jpp} the large~$N$
limit of the ${\cal N}{=}2$ $SU(N)$ gauge theory along the lines of
recent ideas such as those in ref.\cite{juan}.  A large number of
these D6--D2* objects give a supergravity solution, which might be
expected to encode (at least) some of the large~$N$ physics.
Interestingly, the supergravity solution which one naively writes down
suffers from a naked singularity known as a
``repulson''\cite{repulsive} which is unphysical, and incompatible
with the physics of the gauge theory. One expects that there should be
a sensible supergravity solution, valid for $g_sN$ large, where $g_s$
is the string coupling.

In fact\cite{jpp} the repulson is not present, since it represents
supergravity's best attempt to construct a solution with the correct
asymptotic charges. In the solution (not displayed here since we will
not need it; see ref.\cite{jpp}), the volume of the $K3$, set to $V$
asymptotically, actually decreases as one approaches the core of the
configuration. At the centre, the $K3$ radius is zero, and this is the
singularity.

This ignores rather interesting physics, however. At a finite distance
from the putative singularity, the volume of the $K3$ gets to
$V{=}V_*$, so the stringy phenomena ---including new massless
fields--- giving the enhanced $SU(2)$ should have played a role. So
the aspects of the supergravity solution near and inside the special
radius, called the ``enhan\c con radius'', should not be taken
seriously at all, since it ignored this stringy physics.

To a first approximation, the supergravity solution should only be
taken as physical down to the enhan\c con radius $r_{\rm e}$. That
locus of points, a two--sphere $S^{2}$, is itself called an ``enhan\c
con''. It deserves a name, and to be considered as an object in its
own right, since D6--D2* objects probing this geometry seem to spread
out or smear onto it as they approach it, losing their identity (see
ref.\cite{jpp} and later in this paper).  In this way, we see that the
enhan\c con is a hypersurface apparently made of branes which have
puffed up into a sphere. There is a natural generalization of this all
to situations involving branes of different dimensions, (with the
enhan\c con a sphere of different dimensionality), and including
orientifolds.  This pertains to $SU(N)$, $SO(2N)$, $SO(2N{+}1)$ and
$USp(2N)$ gauge theories with eight supercharges in various
dimensions\cite{jpp,jj}.

\subsection{Overview of This Paper}

In the rest of this paper we will uncover many new properties of the
enhan\c con pertaining to the 2+1 dimensional $SU(N)$ gauge theory.
Many of the generic features will have meaning in other dimensions and
for other gauge groups.  We will obtain detailed information because
we can exploit the connection\cite{SWtwo,CH,hanany} to the classical
physics of monopoles. The relevant properties of the enhan\c con
already alluded to so far are reviewed in the next section, and the
details that we will need are emphasized, including the perturbative
expression for the metric on the spacetime geometry as seen by a probe
brane.  Section~\ref{AH} shows that a number of (metric) geometrical
details of the enhan\c con can be learned from the observation that
the full non--perturbative spacetime geometry (as seen by the probe in
the decoupling limit $\alpha^\prime{\to}0$) can be deduced from the
Atiyah--Hitchin manifold\cite{atiyah}, and appropriate generalisations
thereof, which we conjecture to exist as an ADE family.
Section~\ref{mono} focuses on the description of the system of $N$
monopoles {\it via} Nahm data\cite{nahm}.  The point is simply that
the since the description of the $N$ coincident D6--D2* branes
carrying the $SU(N)$ is as classical monopoles, we ought to learn more
about them by studying the well--established technology for describing
multi--monopoles.  In this way, we see that there is some essential
non--commutativity in the description, and we exploit this in
section~\ref{fuzz} to show that the enhan\c con is actually a
``fuzzy'' or non--commutative sphere\cite{fuzzysphere}. This makes
contact with the ``dielectric brane'' construction of
Myers\cite{robdielectric}, and we discuss the similarities and
differences between the two cases.

\begin{figure}[ht]
\centerline{\psfig{figure=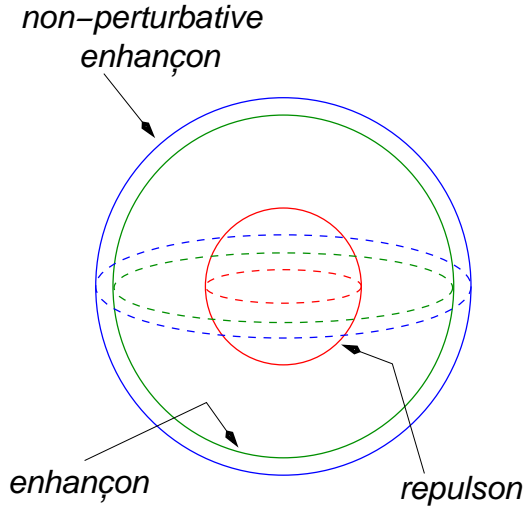,height=2.8in}}
\caption{\small
  A summary of the geometry uncovered. There are three spheres shown:
  The unphysical repulson (innermost) was seen to be removed in
  ref.\cite{jpp}, and replaced by the enhan\c con (next). Here, we
  find that the true non--perturbative enhan\c con radius is slightly
  different from this (shown outermost), although the correction is
  exponentially small at large $N$.  The enhan\c con is a
  non--commutative sphere. There is smooth matching onto the
  spherically symmetric supergravity solution at large $N$.  The
  region interior to the enhan\c con is the core of an
  $(N{-}1)$--monopole.  There is an unbroken $SU(2)$ there.  }
\label{substringy}
\end{figure}

Figure~\ref{substringy} is a summary of some of the properties of the
geometry.

Finally, in section~\ref{candidate} we make a tentative but contentful
conjecture about a possible stringy dual description of the 2+1
dimensional $SU(N)$
gauge theory at large $N$. It is motivated by the fact that the part
of the supergravity geometry which can be written, in the decoupling
limit, entirely in terms of gauge theory quantities (after referral to
the ``frame'' of a monopole probe) is a part of monopole moduli space.
It is suggested that the Nahm data (a family of $N{\times}N$ matrices
depending upon a single coordinate) might be used at large~$N$ to
describe a matrix string theory with many of the properties needed for
a stringy dual. The construction of the string is likely quite
analogous to the more familiar matrix strings, but this new string
theory inherits rich properties of the monopole physics, since it is
built out of Nahm data. Further work is needed on this proposal. We
close with an added note on other work.

\section{$SU(N)$ Gauge theory and BPS Monopoles}
\label{gaugetheory}
In fact, the phenomenology of the enhan\c con in supergravity is
consistent with the monopole physics and with the physics of the 2+1
dimensional $SU(N)$ gauge theory at large~$N$.  The moduli space of
supersymmetric vacua of the theory is parameterized by the vevs of the
three adjoint scalars $\Phi_i$ ($i=3,4,5$) in the vector--multiplet.
This is $3(N{-}1)$ dimensional, since they generically live in the
Cartan subalgebra of the gauge group when satisfying this condition.
At a generic point on this space, the gauge symmetry is therefore
$U(1)^{N-1}$, (hence the name ``Coulomb branch'') and these Abelian
gauge fields may be dualized to give $N{-}1$ more scalars. The moduli
space is therefore $4(N{-}1)$ dimensional. The complete, quantum
corrected moduli space is a smooth hyperk\"{a}hler manifold, given
that there are eight supercharges. In fact\cite{SWtwo,CH,hanany}, the
moduli space\cite{modulispace} is that of $N$ BPS monopoles.

In the probe computation of ref.\cite{jpp}, a single D6--D2* object
was used to probe all of the others, in order to investigate how the
geometry looks from its point of view. As moving the probe slowly in
the background of its siblings is a BPS process, there should be no
potential in the effective Lagrangian for this procedure, and only
kinetic terms. From these terms may be read the metric of spacetime as
seen by the probe\cite{manton,douglas}.  The result of the computation
is:
\begin{equation}
ds^2=
F(r) \left({d r}^2
 +r^2{d\Omega}^2 \right)
+F(r)^{-1}\left({d s}/2-\mu_2C_\phi{d\phi}/2\right)^2\ ,
\label{fourth}
\end{equation}
where
\begin{equation}
F(r)={Z_6\over 2g_s}\left(\mu_6V(r)-\mu_2\right)\ ,
\end{equation}
and ${d\Omega}^2={d\theta}^2+\sin^2\!\theta\,
{d\phi}^2$ and $C_\phi={-}(r_6/g_s)\cos\theta$.
The volume of $K3$ is $V(r)=V Z_2(r)/Z_6(r)$, with
\begin{eqnarray}
Z_2 \* 1+\frac{r_2}{r}\ ,\quad  r_2 =
-\frac{ (2\pi)^4 g_s N \alpha'^{5/2} }{ 2V } \ , \nonumber\\
Z_6 \* 1+\frac{r_6}{r}\ ,\quad  r_6 = \frac{g_sN\alpha'^{1/2}}{2} \ ,
\end{eqnarray}
the harmonic functions appearing in the supergravity solution, which
we do not display here.  The basic D6-- and D2--brane charges
are\cite{polchinski} $\mu_6{=}(2\pi)^{-6}\alpha^{\prime{-7/2}}$ and
$\mu_2{=}(2\pi)^{-2}\alpha^{\prime -3/2}$, respectively.

Notice that the metric (\ref{fourth}) is singular where the monopole's
mass per unit volume, $\tau=(\mu_6V(r){-}\mu_2)/g_s$ vanishes, which is
at
\begin{equation}
V(r)=\mu_2/\mu_6=(2\pi\sqrt{\alpha^\prime})^4{\equiv}V_*\ .
\end{equation} 
This happens at the ``enhan\c con'' radius 
\begin{equation}
r_{\rm e}={2V\over{V} - V_* } |r_2|\ .
\label{enhanced}
\end{equation} This is consistent with the fact that a 
monopole's mass is set by the value of the Higgs, while its size is
inversely proportional to it. So as $\mu$ approaches zero at $r_{\rm
  e}$, a monopole probe becomes smeared out as it merges into all the
other monopoles at the core.  The departure from a sharp description
as a heavy point--like object ---the smearing--- is signaled in the
kinetic energy's divergence.

Since the monopoles cannot go to $r<r_{\rm e}$ in the supergravity
geometry in a supersymmetric way\cite{jpp} consistent with the gauge
dynamics and common sense, it is sensible to conclude that there is
simply new geometry and physics in that region, as anticipated in the
supergravity discussion of the previous section. Perhaps we can learn
more about the enhan\c con by a closer study of the gauge theory, and
hence the monopole physics.

The coupling of the $SU(N)$ gauge theory is given by
\begin{equation}
g_{{\rm YM}}^2 =   (2\pi)^4 g_s\alpha'^{3/2}V^{-1}
\end{equation}
To isolate the gauge theory, it is prudent to focus on the limit where
we attempt a decoupling limit by sending $\alpha^\prime{\to}0$,
holding the coupling and $U{=}r/\alpha'$ finite\cite{juan}. In this
case, the metric becomes
\begin{eqnarray}
&&ds^2=f(U) \left({d U}^2 +U^2{d\Omega}^2\right) +f(U)^{-1}
 \left({d\sigma} -{N\over{8\pi^2}}A_\phi{d\phi}\right)^2\
 ,
\nonumber\\
{\rm where}&&\nonumber\\
&&f(U)={1\over 8\pi^2 g^2_{\rm YM}}
\left(1-{ \lambda \over U}\right)\ ,
\label{probe}
\end{eqnarray}
the $U(1)$ monopole potential is $A_\phi=\pm1-\cos\theta,$ and
$\sigma=s{\alpha^\prime}/2$.  This metric is meaningful only for
$U{>}\lambda$. It is the Euclidean Taub--NUT metric, with a negative
mass.  It is a hyperK\"ahler manifold, because $\nabla
f{=}\nabla{\times}A$, where $A{=}(N/8\pi^2)A_\phi d\phi$.

It is intriguing to note\cite{jpp} (and we shall try to exploit this
more fully later) that only the gauge theory quantities $U$ (a
characteristic energy scale) and $\lambda{\equiv}g_{{\rm YM}}^2 N$
(the 't Hooft coupling) survive the limit, while all other details of
the supergravity have disappeared. The enhan\c con is at
$U{=}\lambda$.  A crucial point which can be read off from this
geometry is that the enhan\c con appears as the one--loop correction
to the gauge coupling, representing the Landau pole. There are
instanton corrections to this (and hence to the manifold), smoothing
out the singular nature at the enhan\c con.  The expectation was
expressed in ref.\cite{jpp} that this manifold is would be thereby
corrected to an Atiyah--Hitchin\cite{atiyah}--like manifold, and we
shall see that this is true presently.

From the point of view of the monopole description, this manifold
should be related to the metric on the moduli space of monopoles. It
is clearly a submanifold of the full $4N{-}4$ dimensional metric on
what is known as the ``strongly centered'' moduli space of $N$ BPS
monopoles\footnote{``Strongly centred'' means that we have the {\sl
    relative} moduli space, where the overall center of mass and
  overall phase of the monopoles are not included.}.

\section{The Role of the Atiyah--Hitchin Manifold}
\label{AH}
Precisely which submanifold we have here should be of interest to us.
First observe that we can change variables in our probe metric
(\ref{probe}) by absorbing a factor of $\lambda/2=g^2_{\rm YM}N/2$
into the radial variable $U$, defining $\rho=2U/\lambda$.  Further
absorb $\psi=\sigma 8\pi^2/N$ and gauge transform to
$A_\phi={-}\cos\theta$. Then we get:
\begin{eqnarray}
&&ds^2= {g^2_{\rm YM}N^2\over 32\pi^2}ds^2_{\rm TN}\ ,
\quad{\rm with}\label{taubnut}\\
&&ds^2_{\rm TN}=\left(1-{2\over\rho}\right) \left({d \rho}^2 
+\rho^2{d\Omega}^2\right) 
+4\left(1-{2\over\rho}\right)^{-1}
\left({d\psi}+\cos\theta{d\phi}\right)^2\ .
\nonumber
\end{eqnarray}
The latter is precisely the form of the Taub--NUT metric that one gets
by expanding the Atiyah--Hitchin metric in large~$\rho$ and neglecting
the exponential corrections.

For the case of $N=2$, the Atiyah--Hitchin manifold is the full
non--perturbative result for the moduli space of the $SU(2)$ gauge
theory\cite{SWtwo}. In fact for this manifold, written in these
coordinates, the periodicity of $\psi$ is $2\pi$ (see later) and so
the $SU(2)$ isometry of the Taub--NUT manifold is broken to
$SO(3)\equiv SU(2)/\Z_2$ by the exponential corrections. The
quotientied sphere $S^3/\Z_2$ at infinity is an orbit under this.

In unscled coordinate of the $U(1)$ probe gauge theory, this means
that the quantity $4\pi^2\sigma={\tilde\sigma}$ is the $2\pi$ periodic
dual scalar to the photon. {\it This periodicity is independent of
  $N$.} Therefore, for arbitrary $N$, the periodicity of $\psi$ is
$4\pi/N$. This allows us to characterise the  manifolds we need for
all~$N$:

They have only {\it local} $SO(3)$ action, and globally the isometry
is broken and we have only the action of $SU(2)/\Z_N$. The manifolds
are therefore asymptotically the negative mass Taub--NUT, and the
space at infinity is an $S^3/\Z_N$. We will characterise the form of
the exponential corrections for arbitrary $N$ in the next section.

\subsection{The Non--Perturbative Corrections}

So this is our first set of information that we learn by studying the
monopole physics: The exponential corrections to our manifold (as seen
by the probe) ---and hence information about the neighbourhood and
interior of the enhan\c con geometry--- are of the same form as those
for the Atiyah--Hitchin manifold (with a generalisation we shall
characterise shortly). This is remarkably fortuitous, and will teach
us more presently.  For definiteness, let us display the full
Atiyah--Hitchin manifold\cite{atiyah,gibbonsmanton}:
\begin{eqnarray}
&&ds^2_{\rm AH}=f^2d\rho^2+a^2\sigma_1^2+
b^2\sigma_2^2+c^2\sigma_3^2\ ,\,\,\,{\rm where}\nonumber \\
&&\qquad\sigma_1=-\sin\psi d\theta+\cos\psi\sin\theta d\phi\ ;\nonumber\\
&&\qquad\sigma_2=\cos\psi d\theta+\sin\psi\sin\theta d\phi\ ;\nonumber\\
&&\qquad\sigma_3=d\psi +\cos\theta d\phi\ ;\nonumber\\
{2bc\over f}{da\over d\rho}&=&(b-c)^2-a^2\ ,\mbox{ and cyclic perms.;}\quad 
\rho =2K\left(\sin{\beta\over2}\right),
\end{eqnarray} and $K(k)$ is the elliptic integral of the first kind:
\begin{equation}
K(k)=\int_0^{\pi\over2}(1-k^2\sin^2\tau)^{1\over2}d\tau\ .
\label{elliptic}
\end{equation}
Also, $k{=}\sin(\beta/2)$, the ``modulus'', runs from $0$ to $1$, so
$\pi\leq\rho\leq\infty$.

The difference between this and negative mass Taub--NUT~(\ref{taubnut})
at large~$\rho$ is exponential, {\it i.e.,} of the form $e^{-\rho}$.
In the case of $SU(2)$ gauge theory ($N=2$), this translates (using
the formulae above~(\ref{taubnut}) into precisely the right form to be
instanton corrections $e^{-U/g^2_{\rm YM}}$, and this has been proven
to be the correct interpretation from a number of points of
view\cite{SWtwo,valya}.

For $SU(N)$, we expect instantons in the field theory to have
essentially the same action, and so this translates into a set of
exponential corrections of the form $e^{-N\rho/2}$. So for large~$N$
therefore, the corrections to the Taub--NUT manifold are quite small,
but the instantons smooth it out on a small enough scale nonetheless.
(This smallness of the intanton corrections to ${\cal N}=2$ $SU(N)$ gauge
theory moduli space at large~$N$ has been noticed in other contexts,
{\it e.g.} in ref.\cite{dougshenk}.)

So in short, our fully corrected moduli space, which also contains
information about the spacetime geometry, is given by a family of
manifolds naturally generalising the Atiyah--Hitchin manifold, after
rescaling $\rho$, and $\psi$. It would be interesting to characterise
these manifolds further. One expects them to be smooth, or at least to
contain $4(N-2)$ parameters which allow them to be deformed to a
neighbouring smooth manifold for the simple reason that the gauge
theory moduli space, not having a Higgs branch to connect to, is
expected to be smooth. There is a concern that this is only true for
the full $4(N-1)$ dimensional manifold representing the moduli space,
and that a restriction to a 4 dimensional submanifold can introduce
singularities. This is where the $4(N-2)$ deformation parameters come
in, as they represent the frozen moduli of the other monopoles/vacua,
now entering as parameters in the reduced theory; intuitively, one
expects the process of a single monopole merging with $N$ others to be
smooth, or at least smoothable by moving the others around.\footnote{I
  am grateful to Gary Gibbons, Juan Maldacena, Robert Myers, and
  Edward Witten for suggestions and comments on this issue.}
 
\subsection{The Case of Two Monopoles}

It is now worth reminding ourselves about the physics of the
Atiyah--Hitchin manifold, using it as a prototype for our case
involving general $N$.  The Atiyah--Hitchin manifold\cite{atiyah} is
the metric on the strongly centred moduli space of two BPS monopoles.
In fact, the two monopole solution itself ({\it i.e.},~the gauge and
Higgs fields) is not spherically symmetric\footnote{This is generally
  true for the $N$ monopole solution\cite{bogo}, as we shall
  discuss.}. At best, it is axisymmetric, and this is when the two
monopoles are coincident. The coordinate $\rho$ represents the
asymptotic separation of the monopoles. It really only has this
meaning when the monopoles are separated quite far apart, and then the
metric reduces to $ds^2_{\rm TN}$. Closer than this, the monopoles
cease to be distinct.  Actually, the singularity at $\rho=2$ in the
Taub--NUT metric is completely meaningless, as it is well outside the
range of validity of the large~$\rho$ expansion used to get that
metric. Further proof of this comes from the fact that the monopoles
are coincident at $\rho=\pi$. This special (axisymmetric)
solution\cite{ward,manymono} has monopole charge 2, and really has no
sensible description in terms of individual monopoles at all.
Generically, all we can say (this can be confirmed by a study of the
location of the zeros of the Higgs field part of the monopole
solution\cite{brown}) is that for any $\rho$ the monopoles are spaced
symmetrically along an axis, which we can choose to be the $x_3$ axis.
The Higgs field has zeros even when the monopoles cease to have any
sensible meaning (since they grow large and diffuse), and are often
used as a guide to the ``location'' of the monopoles, despite their
finite coresize.  When $\rho=\pi$, the two Higgs zeros are both at the
origin, and this is the coincident case.

Note that despite the fact that the solution is axisymmetric, far away
from it, in spacetime ($r\to\infty$), the Higgs field is
\begin{equation}
{1\over2}\Tr[\H(r)^*\H(r)]=H-{Ne_{\rm m}\over r}+\cdots
\label{higgs}
\end{equation}
for (here) $N=2$, and this form of the Higgs is generally true for all
$N$. Here $e_{\rm m}=2\pi/{e}$ is the basic unit of magnetic charge,
Dirac fixed in terms of the electric charge $e$.

Actually, the metric components in the neighbourhood of $\rho=\pi$
are\cite{gibbonsmanton}:
\begin{eqnarray}
&&a=2(\rho-\pi)+O\left((\rho-\pi)^2\right)+\cdots\ ,\nonumber\\
&&b=\pi+O\left((\rho-\pi)^2\right)+\cdots\ ,\quad 
c=-\pi+O\left((\rho-\pi)^2\right)+\cdots
\end{eqnarray}
and so the metric appears to be singular there, given that $a{\to}0$.
In fact, the $S^3$ of $(\psi,\theta,\phi)$ collapses to a two--sphere,
$S^2$, there, but this point is actually a coordinate ``bolt''
singularity.  It is the removal of this bolt which requires $\psi$ to
be $2\pi$ periodic, a fact which featured in the previous two
subsections.

It should be noted that we have only described a simple cover of the
moduli space of two monopoles. There is an addition $\Z_2$ which
identifies configurations which correspond to each other after an
exchange of the (identical) monopoles. In this way the bolt becomes an
${\rm RP}_2$ instead of an $S^2$. We will not have such a symmetry
here, so our bolt (or generalisation thereof), for $N>2$ will always
be an~$S^2$.

\subsection{Large $N$ and Spacetime Physics}
How are we to make sense of the appearance of the
Atiyah--Hitchin--like manifolds in our case, and what can we learn
about spacetime physics?  Well, we have a four dimensional submanifold
of the full metric on moduli space, and so most of the parameters
($4N{-}8$ of them) have been fixed. In ref.\cite{jpp} all of the
branes were placed at the origin $r=0$ in the supergravity discussion.
So all of the parameters were frozen except the single probe brane's
position and phase.  Although in supergravity they are naively at
$r=0$, this is not the case, and they are smeared out into a sphere of
radius $r_{\rm e}\sim N$.

Now, the Atiyah--Hitchin coordinate $\rho$ should have an
interpretation as a separation from the center of mass. For the
2--monopole case that would tell us very little about the spacetime
geometry, but since we have $N$ monopoles, and $N$ is large,
$U=g^2_{\rm YM}\rho N/2$ is a good radial coordinate for spacetime, as
the center of mass is still close to $U=0$ when only one probe is
separated off.  {\sl This is why at large~$N$ our scaled
  Atiyah--Hitchin--like manifold has a dual meaning as a relative
  moduli space for monopoles as well as the spacetime geometry seen by
  the probe.}  For small $N$, the coordinate $U$ is not as good a
guide to the spacetime geometry.

Note that for any $N$, the spacetime Higgs field will asymptotically
behave as in equation~(\ref{higgs}).  This is why the supergravity
solution can be spherically symmetric, as its asymptotically
spherically symmetric geometry matches on to this behaviour, while the
deviation from spherical symmetry is a detail only visible in terms
subleading in large~$N$. Taking the expression for the volume $V(r)$
and expanding gives:
\begin{equation}
{V(r)\over V_*}-1=\left({V\over V_*}-1\right)
-\left({V\over V_*}+1\right){g_s\alpha^{\prime 1/2}N\over r}+\cdots\ ,
\end{equation}
confirming the earlier statement about the relation between the volume
and the Higgs field, and fixing $H=(V/V_*{-}1)$ and $e_{\rm
  m}{=}(1+V/V_*)g_s\alpha^{\prime 1/2}/2$.  (Later, we will set $H=1$
and hence $V=2V_*$, since we are free to choose these parameters at
our convenience.  Note that the explicit one--monopole solution
displayed in eqn.(\ref{one}) is so normalized, and has $e_{\rm m}$ set
to 1.)

Taking seriously the other lessons learned from the two monopole case,
the analogue of the bolt sphere at $\rho=\pi$ is where the probe
merges into all the other branes. For small $N$, the correction from
$\rho=2$ to the bolt radius is significant, while for large~$N$, as we
have seen, the instanton corrections are small. Inside the bolt radius
there is really no meaning to the coordinate $\rho$ as having anything
to do with distinct monopoles. In fact, this is very robust:
Scattering two identical monopoles using the Atiyah--Hitchin manifold
shows that $\rho=\pi$ is truly the distance of closest approach: a
head--on collision results in a $90^{\rm o}$ scattering angle at
$\rho=\pi$.  The finite core size of the monopoles takes over.  We
inherit this qualitative behaviour here for arbitrary $N$.

So in fact we learn that there is indeed a sharp meaning to the sphere
of closest approach for the monopoles. It is also where they become
massless, and also become indistinct. It is precisely where there
occurs a bolt {\it coordinate} singularity in the {\it smooth}
Atiyah--Hitchin manifold.  The radii $\rho<2$ (or $U<g^2_{\rm YM}N$)
do not have any meaning for the individual monopole probes.

\section{The Multi--Monopole from Nahm Data}
\label{mono}
In brane/supergravity language, we naively placed the branes all at
the same place ($r=0$), at the point of $SU(N)$ symmetry, the origin
of the Coulomb branch. As we know from other
examples\cite{seibergwitten}, quantum corrections in the gauge theory
alter the structure of that point. In fact, the enhan\c con is
precisely a manifestation of this, since the monopoles are really not
at the origin, but smeared into a sphere.

One of the crucial points of the present investigation is that the
entire physics of the moduli space of the gauge theory is given in
terms of the {\sl classical} BPS monopoles, and so we should look no
further than that system in order to learn more about the enhan\c con,
and what it means.  So how does one describe a clump of $N$ monopoles?

Charge $N$ multi--monopoles generalising the charge 1 BPS
solution\cite{onemonopole} were constructed by a number of elegant
techniques.  On the one hand, algebraic techniques\cite{atiyahward}
were employed by Ward\cite{ward}, with
generalisations\cite{nmonopoles}; on the other hand, a
connection\cite{backlund} to B\"acklund transformations and inverse
scattering techniques was employed in refs.\cite{manymono,forgacs}.
Since the Bogomolnyi equations (\ref{bogomolnyi}) are related by
dimensional reduction to the self--dual equations in four Euclidean
dimensions, there is another elegant description {\it via} an
extension of the ADHM construction\cite{nahm,adhm,donaldson}.  The
basic (``covariant'' Nahm) equations are:
\begin{equation}
{d\Phi^i\over d\sigma}+[\Phi_0,\Phi_i]=
{1\over 2}\epsilon_{ijk}[\Phi^j,\Phi^k]\ ,
\label{nahm}
\end{equation}
where $i,j,k$ run over $1,2,3$. Here, $\Phi_0(\sigma)$ and
$\Phi_i(\sigma)$ are $N{\times}N$ anti--Hermitian $SU(N)$ matrices,
with $\Phi^*_i(\sigma)=-\Phi_i(\sigma)$ and $\Phi_i(\sigma)=-{\bar
  \Phi}_i(-\sigma)$. The coordinate $\sigma$ has range
$-H\leq\sigma\leq H$, where $H$ is the asymptotic value of the Higgs
field, which we shall presently set to 1 in much of the rest of this
paper.  The data appropriate to monopoles arise as solutions to this
equation which are regular in the interior of $[-H,H]$, with
appropriate boundary conditions at $\sigma=\pm H$. Those boundary
conditions require that $\Phi_i$ have simple poles there, and that the
residues of those poles (which are of course $N{\times}N$ matrices)
are irreducible representations of $SU(2)$.  There is an $SU(N)$ gauge
invariance,
\begin{equation}
\Phi_0\to G\Phi_0G^{-1}-{dG\over d\sigma}G^{-1}\ ,
\qquad \Phi_i\to G \Phi_i G^{-1}\ ,
\label{gauge}
\end{equation}
where the $G(\sigma)\in SU(N)$, and are the identity at the ends of
the interval.  There is a specific construction (also following the
ADHM techniques) for converting the solutions of these equations
---the ``Nahm data''--- into expressions for the spacetime fields
$A_i({\bf x})$, $\H({\bf x})$, which we shall not reproduce here since
for $N>1$, closed forms are not known.  Hitchin\cite{hitchin} has
shown using algebraic methods that this method constructs all of the
monopole solutions and indeed that it is equivalent to the
aforementioned monopole constructions based on those of
Ward\cite{ward,nmonopoles}.

The gauge transformations (\ref{gauge}) can be used to set $\Phi_0$ to
zero, giving the standard Nahm equations, but should be left unfixed
in order to perform the full hyperK\"ahler
quotient\cite{donaldson,adhm} which constructs the metric on the
moduli space of Nahm data. Nakajima\cite{nakajima} has shown that the
metric thus computed is indeed the monopole moduli space, and it is smooth.

This Nahm system arises naturally in the brane description as the
condition on the brane fields for supersymmetric vacua, resulting in a
hyperk\"{a}hler quotient.  The $\Phi$ are adjoint scalars in a gauge
theory on the brane. The most natural brane system where this arises
is probably that of\cite{diaconescu} $N$ D1--branes stretched
perpendicularly between two D3--branes separated by a distance~$2H$.
The coordinate $\sigma$ is that along the D1--branes, and the $\Phi$
are the positions of the D1--branes inside the D3--branes. The
boundary conditions arise by considering the 1+1 dimensional theory on
the world--volume as a theory with ``impurities'' located at
$\sigma=\pm H$, which is natural, since the massless 1--3 strings are
localized there\cite{tsimpis,sethi}. These Nahm equations can also be
derived in the brane wrapped on $K3$ system we started with
here\footnote{This follows since these systems are dual to one
  another\cite{jpp}.  The details will appear in ref.\cite{jj2}}. The
asymptotic value of $K3$, $V$, sets the parameter $H$ {\it via}
$H=(V/V_*{-}1)$. As the supergravity parameter $r$ runs from $\infty$
to $r_{\rm e}$ (or more properly as $U$ runs from $\infty$ to
$\lambda$) the coordinate $\sigma$ runs from $H$ to zero. Let us set
$H=1$ henceforth.

\section{The Enhan\c con as a Fuzzy Sphere}
\label{fuzz}
It is easy to see that the enhan\c con is itself a fuzzy sphere in
{\sl spacetime} as follows. The D6--D2* system is dual to a system of
$N$ D3--branes stretched between a pair of NS5--branes in type~IIB
string theory. The $SU(N)$ gauge theory is on the flat part of the
D3--branes. The Nahm equations above (\ref{nahm}) have the following
meaning: The $\Phi_i(\sigma)$, multiplied by
$2\pi\sqrt{\alpha^\prime}$, are coordinates in the $\IR^3$ part of the
NS5--branes where the D3--branes end.  The coordinate $\sigma$ is the
coordinate between the NS5--brane. The 5+1 dimensional theory on the
NS5--branes is the spontaneously broken 3+1 dimensional $SU(2)$ theory
if we ignore the two spatial directions common to both the
branes\cite{hanany}. Translating further, there is a factor of ${1/g_s}$
in front of the commutator in the Nahm equation. (This is instead of
$g_s$, appropriate to the case of D1--branes ending on D3--branes.)

\begin{figure}[ht]
  \centerline{\psfig{figure=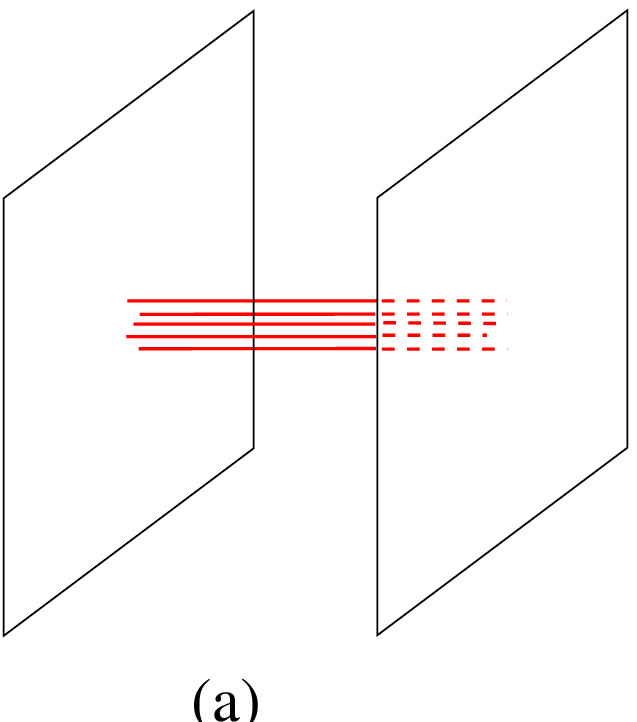,height=2.0in}
    \hskip2cm\psfig{figure=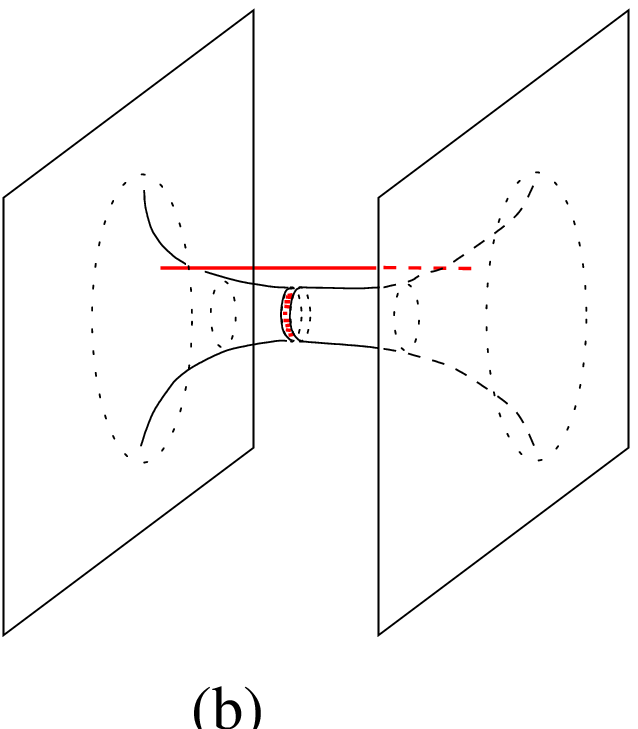,height=2.1in}}
\caption{\small
  (a) The configuration of D3--branes stretching between NS5--branes.
  (b) The resulting ``double trumpet'' shape of the NS5--branes at
  large~$N$. (A separated probe is also shown.)  This system has a
  natural description in terms of the Nahm equations as explained in
  the text. The enhan\c con is the place (an $S^2$) where the
  NS5--branes touch.}
\label{trumpet}
\end{figure}

There is a ``double trumpet'' shape describing the $N$ stretched
D3--branes pulling on the NS5--branes, as depicted in ref.\cite{jpp}
and reproduced in figure~\ref{trumpet}.
At the centre of the shape, there is a two--sphere where the
fivebranes touch, restoring the $SU(2)$.  Crucially, this is only a
two--sphere for $N$ large enough, since the radius of the sphere is
proportional to $N$, and only for spheres large enough are we far
enough away from the details of the interior of the multi--monopole
configuration to see an approximately spherically symmetric situation.
(Recall that the multi--monopole is not spherically
symmetric\cite{bogo}.)

We can make this a bit more precise as follows: First note that in the
one--monopole case, the solution is spherically symmetric, and the
Nahm data is simply the one dimensional representation of $SU(2)$,
{\it i.e.,} all the $f_i(\sigma)$ are equal.  Assume that in our case,
we are far away enough from the core that we can borrow this
behaviour, making the symmetric choice $\Phi_i(\sigma)=-{\rm
  i}f(\sigma)\Sigma_i$. In doing this, we connect to the discussion of
ref.\cite{robfunnel}. There, this was shown to correspond to an
infinite trumpet shape representing $N$ D1--branes merging into an
orthogonal D3--brane. The required poles at the ends of the Nahm
interval correspond to the flaring of the trumpet as it expands into
the perpendicular shape.  The Nahm equations become:
\begin{equation}
{df\over d\sigma}=
-{f^2\over g_s}\ ,
\end{equation}
and the solution we seek here is
made by gluing together two copies of the trumpet end to end at the
centre of the interval:
\begin{equation}
\quad f(\sigma)=
{g_s\over\sigma\mp 1}\ ,\quad \mbox{ for $\sigma\in[0,\pm1]$}\ .
\end{equation}
This gives a shape like that in figure \ref{trumpet}, but is only an
approximation. The full solution should connect smoothly through the
interior of the interval.  A cross section at some value of $\sigma$
is a non--commutative, or ``fuzzy'' sphere\cite{fuzzysphere} of radius
given by (remembering to put in the factor of
$2\pi\sqrt{\alpha^\prime}$ for dimensions)
\begin{equation}
R^2={4\pi^2\alpha^\prime}
\sum_i {\rm Tr}(\Phi_i^2)={4\pi^2\alpha^\prime}(N^2-1)f^2(\sigma)\ ,
\end{equation}
There is a minimum value, $f_{\rm e}\sim g_s$ where the NS5--branes
touch at $\sigma=0$.  There, the radius is
\begin{equation}
R_N=2\pi\sqrt{\alpha^\prime} g_s\sqrt{N^2-1}\sim 
2\pi \sqrt{\alpha^\prime} g_sN\ ,
\end{equation}
which compares well with the supergravity expression (\ref{enhanced})
for the enhan\c con radius, which is
\begin{equation}
r_{\rm e}=g_sN \sqrt{\alpha^\prime}\left({V\over V_*}-1\right)^{-1}\sim
{N\over He}\ .
\end{equation}
This particular fuzzy sphere is the enhan\c con. It is a sensible
smooth sphere of non--zero radius at large~$N$. Notice also that it
has roughly the correct behaviour for the $N$ monopole size in terms
of the Higgs vev $H$ and the electric charge $e$.

\begin{figure}[ht]
\centerline{\psfig{figure=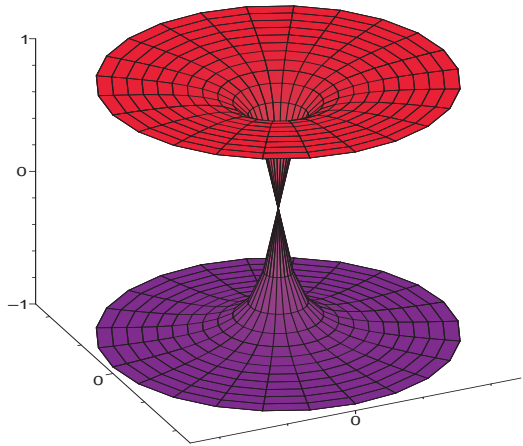,height=2.3in,width=2.7in}}
\caption{\small
  A plot of a slice of the two NS5--branes' shape, as deduced from the
  Higgs field (\ref{one}) for the single monopole. It is the only
  exactly spherically symmetric case. The double trumpet shape pinches
  off to zero size in this case.}
\label{single}
\end{figure}

For small $N$ ($\neq1$) it is very non--spherical, while it collapses
to zero size in the case $N{=}1$: The minimum value $f_{\rm e}$ is
zero for a single monopole, and the double trumpet profile pinches
off, as can be deduced (see ref.\cite{aki}) from a study of the Higgs
field (\ref{one}) for the explicitly known one--monopole solution. We
plot this in figure~\ref{single}.

Returning to large~$N$, it is in this sense that we see the connection
between the dielectric brane construction of ref.\cite{robdielectric}
(see also ref.\cite{kabatrey}) (where branes puff up into a sphere in
the presence of a background R--R field) and the enhan\c
con\footnote{Non--commutativity in the enhan\c con geometry was
  suspected in ref.\cite{jpp}, and a relation to the dielectric branes
  was suspected by many.}.  Both phenomena can be used as examples of
a new mechanism for excising undesirable spacetime
singularities\cite{singclass}, but the dielectric mechanism is adapted
to ${\cal N}{=}1$ supersymmetry preserving vacua\cite{joematt}, while
here we have ${\cal N}{=}2$ (counting in four dimensional units). The
connection between the two is simply that there are non--zero
commutators for the adjoint scalars (forming~$N$--dimensional
irreducible representations of $SU(2)$) which have been shown in one
case\cite{robdielectric} to induce multipole couplings to higher rank
R--R fields. This is equivalent to the growth of extra dimensions on
the brane.

In the present case, the Nahm equations are the means by which
non--zero commutators arise in an ${\cal N}{=}2$ supersymmetry
preserving way.  The higher dimensional aspect of the branes is
realized in terms of their description as a finite sized
multi--monopole configuration. The non--commutative sphere which is
the enhan\c con is a preferred slice through this geometry, forming
the effective shell around the $N$--monopole core.

\section{A Candidate Dual?}
\label{candidate}
One of the goals of a study of the supergravity solution mentioned in
the introduction was to see if there is a large~$N$ dual supergravity
solution.  As pointed out in ref.\cite{jpp}, even with the improved
understanding of the geometry by recognizing the role of the enhan\c
con, the decoupling/scaling limit $\alpha^\prime{\to}0$ (with
$U=r/\alpha^\prime$ finite) gives a ten dimensional supergravity
solution which was not appropriate as a truly decoupled dual theory.
One sign of this (among others) is the simple fact that the resulting
geometry contained parameters which recalled the original data of the
type~IIA compactification.  In other words, it did not assemble into
an expression referring purely to gauge theory quantities, as happens
in simpler cases where there is a genuine supergravity
dual\cite{juan}.

It would certainly be an excellent situation if there was a large~$N$
dual theory all the same, and a persistent open question is whether
there exists such a theory, and whether it is a {\sl useful} dual, in
the sense of being weakly coupled (or at least tractable) when the
gauge theory is strongly coupled.

To this end, let us note again that in the decoupling limit, the part
of the supergravity geometry transverse to the branes assembles into
purely gauge theory quantities, when it is referred to from the
``frame'' of a brane probe.  We ought to regard this as a clue. As the
geometry is (a subspace of) the moduli space of multi--monopoles, this
leads one to speculate that there may be something to be gained by
focusing one's attention there.

On general grounds, one might expect that this large~$N$ dual theory
might be a string theory whose world sheet genus expansion is
isomorphic to the $1/N$ expansion in the planar diagrams of the field
theory, in the usual manner\cite{thooft}.  Besides this, the putative
string theory should encode the structure of (and allow access to) the
$4N{-}4$ dimensional moduli space of the Coulomb branch of vacua.  The
$SU(2)_R$ symmetry of the gauge theory should be present, as should
the phenomenon of the enhan\c con, {\it etc}.

Here is a proposal for such a string theory.  Simply take the large~$N$
limit of the Nahm data, but looked at in a different way!  The
$N{\times}N$ matrices $\Phi_i(\sigma)$, for $(i=0,1,2,3)$, which
satisfy the Nahm equations, may be thought of at large~$N$ as giving
the coordinates of a string in a {\sl four} dimensional transverse
space. It is a matrix string constructed by letting the matrices
explore the full $4N{-}4$ dimensional moduli space (or possibly a
cover of it), which is imposed by the Nahm equations (\ref{nahm}) and
the accompanying gauge invariance (\ref{gauge}).

This proposal is more easily motivated by analogy with a simpler
matrix string construction, that of the ten dimensional case of
type~IIA. There, we have eight $N{\times}N$ matrices $X^i(\sigma)$.
While they may be thought of the collective coordinates of $N$
D1--branes, and hence parameterising $(\IR^8)^N/S_N$, the
now--standard route\cite{motl,dvv} reinterprets them at large~$N$ as
light cone ``string fields'' giving a description the shape of a
single IIA string in eight transverse dimensions $\IR^8$. The
``second--quantized'' description of the string theory is simply the
(1+1)--dimensional $U(N)$ gauge theory of the $X^i$.

In fact, the very same structures are present here, but yielding a
(with respect) potentially much more interesting string, at least for
our purposes, since it contains all of the ingredients to make a dual
string for the (2+1)--dimensional $SU(N)$ gauge theory.  The naive
interpretation of the $\Phi_i(\sigma)$'s is as the non--commutative
collective coordinates of $N$ D--strings stretched along a finite
interval and acting as BPS monopoles. (We are ignoring the two
directions common to both types of brane in the setup of
section~\ref{fuzz}.)  The proposal here is that at large~$N$, the Nahm
data $\Phi_i(\sigma)$ can be thought of as string fields for a single
string having a 4 dimensional transverse target space. To obtain the
``long string'' sectors required to enable a stringy limit, two
modifications to our setup can be considered: The first is that the
whole system of NS5--branes described in section~\ref{fuzz} be placed
on a circle, by compactifying the direction in which the branes are
separated.  The second is that the Donaldson--Nahm $U(N)$ gauge
transformations~(\ref{gauge}), which reduce to the identity at
$\sigma=\pm1$ \cite{donaldson}, be allowed to include permutations at
the ends of the $\sigma$ interval in this periodic system. In this
way, we include configurations in the theory corresponding to a
D--string winding around a large number of times before terminating on
an NS5--brane.

A periodic version of the 1+1 dimensional ``impurity'' gauges theories
of the sort described in refs.\cite{tsimpis,sethi} should provide the
dynamics. To get to strong coupling for the gauge theory, one must
further tune the length of the interval (and hence the Higgs vev) and
the size of the circle to be small, holding the ratio fixed.

By construction, the string thus defined has many of the properties
which we seek for our dual string.  It refers correctly to the moduli
space of vacua of the gauge theory by using the monopole moduli space
in an essential way.  Different vacua of the gauge theory correspond
to different backgrounds for the string theory.  

A tantalizing consequence of this conjecture (which clearly needs more
work) for a dual string is that it may provide a dictionary between
many of the elegant results about monopole moduli space (such as the
scattering of slowly moving monopoles, geodesics representing bound
states, {\it etc.}), and properties of the dual gauge theory.  It will
certainly be interesting to pursue this further.

\section*{\bf Added Note on Other Work}

There is other recent work on the large~$N$ limit of monopoles and
Nahm data, with connections to brane configurations, and
non--commutativity\footnote{The author is grateful to Micha Berkooz
  and Kimyeoug Lee for pointing out refs.\cite{klee,berkooz} after
  this manuscript first appeared.} which we ought to mention here.
While they appear to be separate avenues of investigation, it should
be interesting to learn if there is anything in those approaches which
might help with some of the issues discussed here.

Existing proposals by Fairlie and collaborators\cite{david} concerning
a matrix/M--theory interpretation of Nahm equations (in diverse
dimensions) at large~$N$ emphasize a connection to the Moyal
bracket(see also ref.\cite{infiniteward,linda}), to non--commutative
geometry, and the physics of membranes.  In ref.\cite{klee}, Lee
studies the case of an infinite sheet of BPS monopoles, highlighting
its non--commutative description, and a relation to D1--strings
stretching between D3--branes.

It is interesting to note that both of these sets of work make
connections at large~$N$ to a natural two dimensional non--commutative
surface, while in this paper, we have pointed out that the enhan\c con
itself is a non--commutative sphere. Perhaps a firmer connection might
be made between these approaches which might lead to a description of
the dynamics of the enhan\c con as a non--commutative membrane {\it
  via} those techniques.

Also, the work of Berkooz\cite{berkooz} studies a 1+1 dimensional
impurity model (related to the one we have in mind in this paper) as a
DLCQ realization of the 4+1 dimensional $SU(N)$ Yang--Mills theory
found on D4--branes.  Features such as the crossover between open and
closed string effective descriptions of the gauge theory at large~$N$
are highlighted.


\section*{Acknowledgements}
I am grateful to Peter Forgacs for pointing out my unintentional
omission (in release 1 of this paper) of the excellent independent
early work\cite{manymono,forgacs} on multi--monopoles.  I would like
to thank Alex Buchel, David Fairlie, Gary Gibbons, Juan Maldacena, Rob
Myers, Amanda Peet, Joe Polchinski, Eric Weinberg and Edward Witten
for useful comments, and Samantha Butler for her patience.

\bigskip
\bigskip
\centerline{\psfig{figure=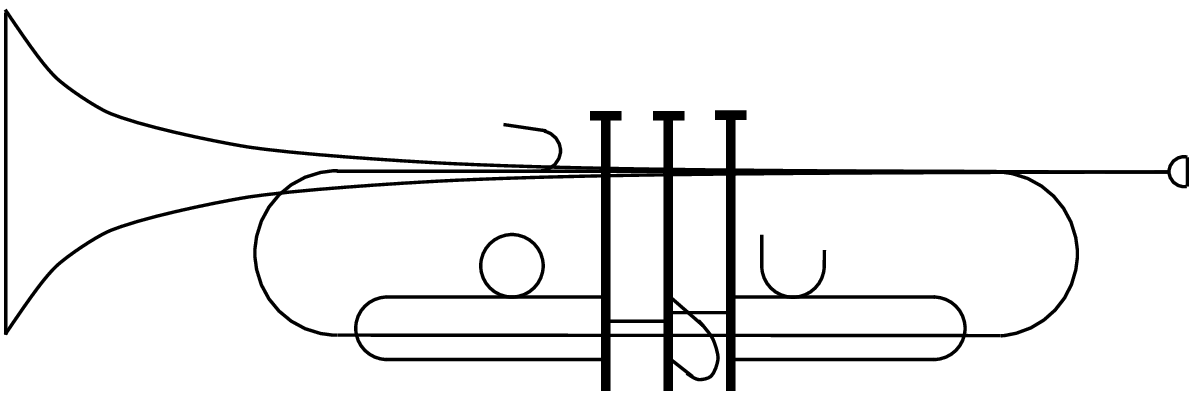,height=0.45in}}

\end{document}